\documentclass[10pt]{article}
\usepackage{spconf,amsmath,graphicx,epsfig,multirow,adjustbox,booktabs,caption,color,amssymb}
\usepackage[colorlinks,linkcolor=red]{hyperref}
\usepackage{lipsum}
\usepackage{fancyhdr}
\usepackage{threeparttable}
\usepackage{float}
\usepackage{stfloats}

\newcommand\blfootnote[1]{%
\begingroup 
\renewcommand\thefootnote{}\footnote{#1}%
\addtocounter{footnote}{-1}%
\endgroup 
}

\title{CS-Rep: making speaker verification networks embracing re-parameterization}

\name{Ruiteng Zhang$^1$, Jianguo Wei$^{1,2}$, Wenhuan Lu$^1$, Lin Zhang$^{3}$, Yantao Ji$^{4}$, Junhai Xu$^{1}$, Xugang Lu$^{5}$ }
\address{$^1$College of Intelligence and Computing, Tianjin University, Tianjin, China\\
	$^2$Computer College, Qinghai Nationalities University, Xining, China\\
	$^3$National Institute of Informatics, Tokyo, Japan\\
	$^4$School of Software Engineering, Xi’an Jiaotong University, Xi’an, China\\
	$^5$National Institute of Information and Communications Technology, Kyoto, Japan}
\begin{document}
\ninept
\maketitle
\begin{abstract}
Automatic speaker verification (ASV) systems, which determine whether two speeches are from the same speaker, mainly focus on verification accuracy while ignoring inference speed. 
However, in real applications, both inference speed and verification accuracy are essential. 
This study proposes cross-sequential re-parameterization (CS-Rep), a novel topology re-parameterization strategy for multi-type networks, to increase the inference speed and verification accuracy of models. 
CS-Rep solves the problem that existing re-parameterization methods are not suitable for typical ASV backbones.
When a model applies CS-Rep, the training-period network utilizes a multi-branch topology to capture speaker information, whereas the inference-period model converts to a time-delay neural network (TDNN)-like plain backbone with stacked TDNN layers to achieve the fast inference speed. 
Based on CS-Rep, an improved TDNN with friendly test and deployment called Rep-TDNN is proposed. 
Compared with the state-of-the-art model ECAPA-TDNN, Rep-TDNN increases the actual inference speed by about 50\% and reduces the EER by 10\%. 
\end{abstract}
\begin{keywords}
Speaker verification, cross-sequential transformation, re-parameterization, inference speed
\end{keywords}
\section{Introduction}
\label{sec:intro}
Automatic speaker verification (ASV) systems determine whether two speeches are from the same speaker or not. The front-end models of ASV systems to extract speaker embeddings from speech utterances to represent the corresponding speaker, such as i-vector \cite{I-vector}, d-vector \cite{d-vector}, and x-vector \cite{x-vector}. The back-end functions of ASV systems calculate similarity scores between two embeddings, e.g., probabilistic linear discriminant analysis(PLDA)\cite{plda} and cosine similarity. 
\blfootnote{Junhai Xu is the corresponding author. Thanks to NSFC of China (No.61876131, No. U1936102), Key R\&D Program of Tianjin (No.19ZXZNGX00030). Regional Innovation Cooperation Project of Sichuan (Grant No.22QYCX0082)}

Currently, ASV systems are widely used in many real-world applications, such as criminal investigation, payment \cite{XU2020394}, and real-time meeting recordings \cite{horiguchi2020end}. In those realistic scenarios, the inference speed is becoming increasingly important.

Recently, the use of multi-branch topology (e.g., ResNet \cite{resnet} has a convolutional branch and a shortcut branch) \cite{xie2017aggregated, gao2019res2net} to improve the theoretical efficiency and accuracy of models has been reported. These designs capture multi-scale information through multi-type convolution kernels with a small number of channels. Additionally, many studies \cite{ecapa-tdnn, zhang21_interspeech} have used multi-branch designs to strengthen the time-delay neural network \cite{x-vector}(TDNN). 
ECAPA-TDNN\cite{ecapa-tdnn} with multi-branch topology has exhibited excellent performance in recent ASV competitions \cite{nagrani2020voxsrc,voxsrc_ecapa_1, voxsrc_ecapa_2, zeinali2019short,thienpondt2021integrating}. Unfortunately, the actual inference speed of the multi-branch designs is typically much slower than its theoretical speed.
The main reason is that the multi-branch designs increase the memory access cost (MAC) and are unfriendly for parallel computing \cite{ma2018shufflenet}. Therefore, this topology does not solve the contradiction of speed and performance in the real applications.

To improve the actual inference speed for the multi-branch models, Ding et al. \cite{ding2021repvgg} proposed a re-parameterization approach to convert the multi-branch network to a plain topology in the inference-period. However, their strategy is based on the structure of ``conv-bn-activation,” but the mainstream TDNN-based ASV models rely on the ``conv-activation-bn” structure to realize good verification accuracy (we confirm that this is reasonable in Section \ref{sec:exp_results}). 
Many successful implementations of ASV have adopted the structure, such as the x-vector \cite{x-vector} (implemented by Kaldi \cite{povey2011kaldi}), E-TDNN \cite{garcia2020jhu}, and ECAPA-TDNN \cite{ecapa-tdnn}, etc. 
In those ASV models, since their convolutional layer is not adjacent to the batch normalization (BN) \cite{ioffe2015batch} layer, they cannot be re-weighted through Ding's method.

\begin{figure*}[h]
\setlength{\abovecaptionskip}{-1pt}
\setlength{\belowcaptionskip}{-2pt}
\begin{minipage}[t]{0.24\linewidth}
  \centering
  \centerline{\epsfig{figure=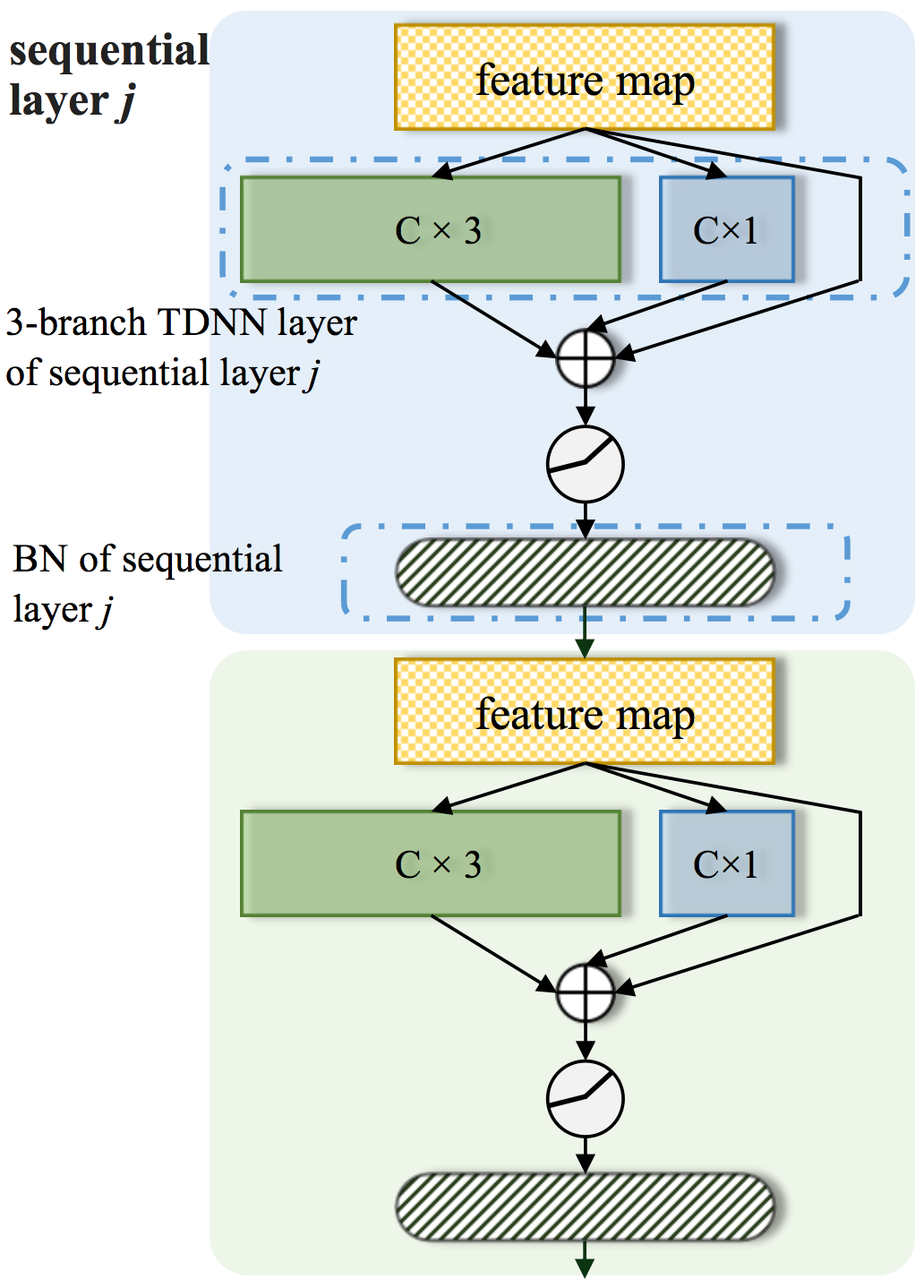, height=5cm}}
  \centerline{{(a)}}\medskip
\end{minipage}
\begin{minipage}[t]{0.26\linewidth}
  \centering
  \centerline{\epsfig{figure=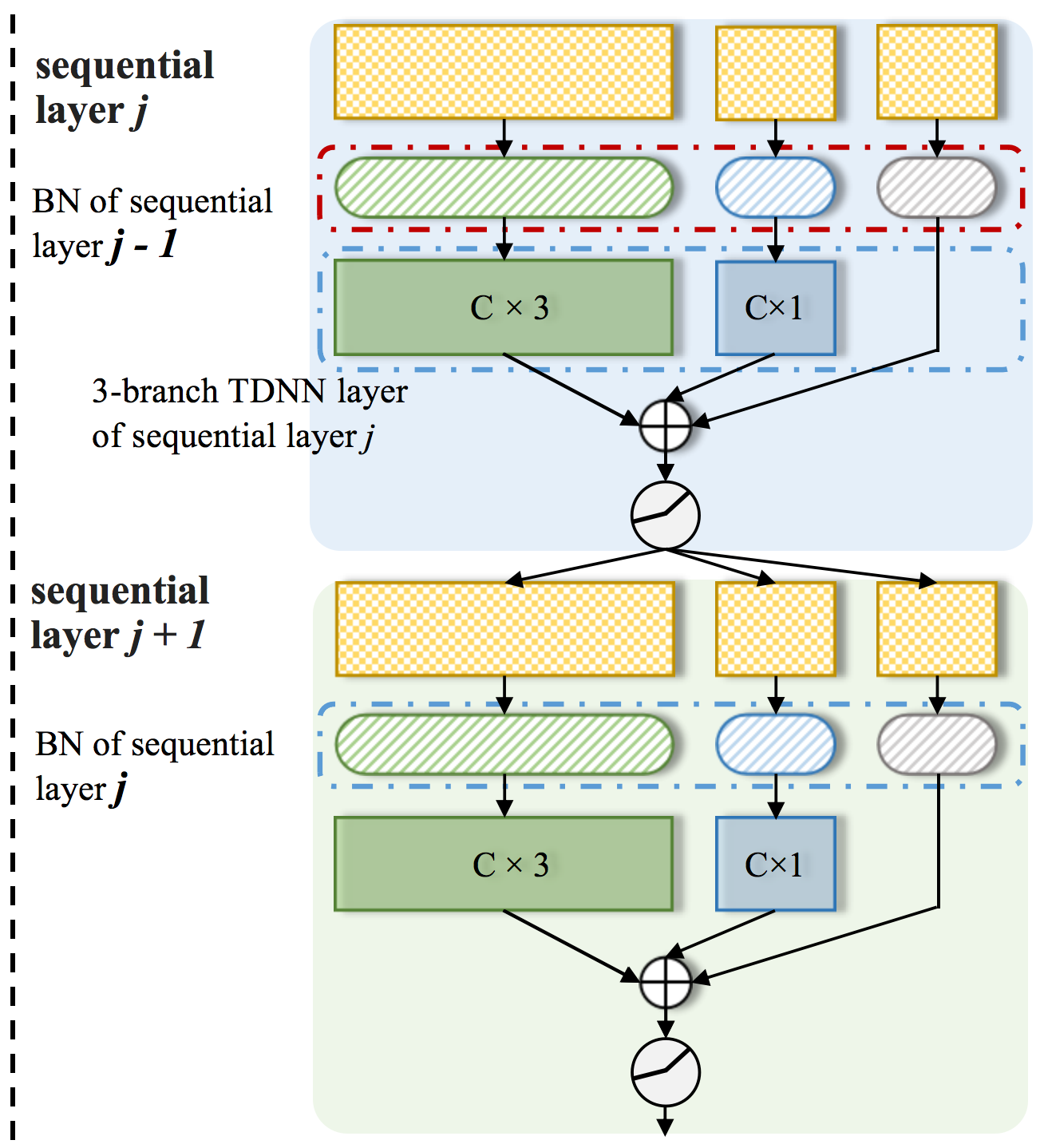, height=5cm}}
  \centerline{{(b)}}\medskip
\end{minipage}
\begin{minipage}[t]{0.24\linewidth}
  \centering
  \centerline{\epsfig{figure=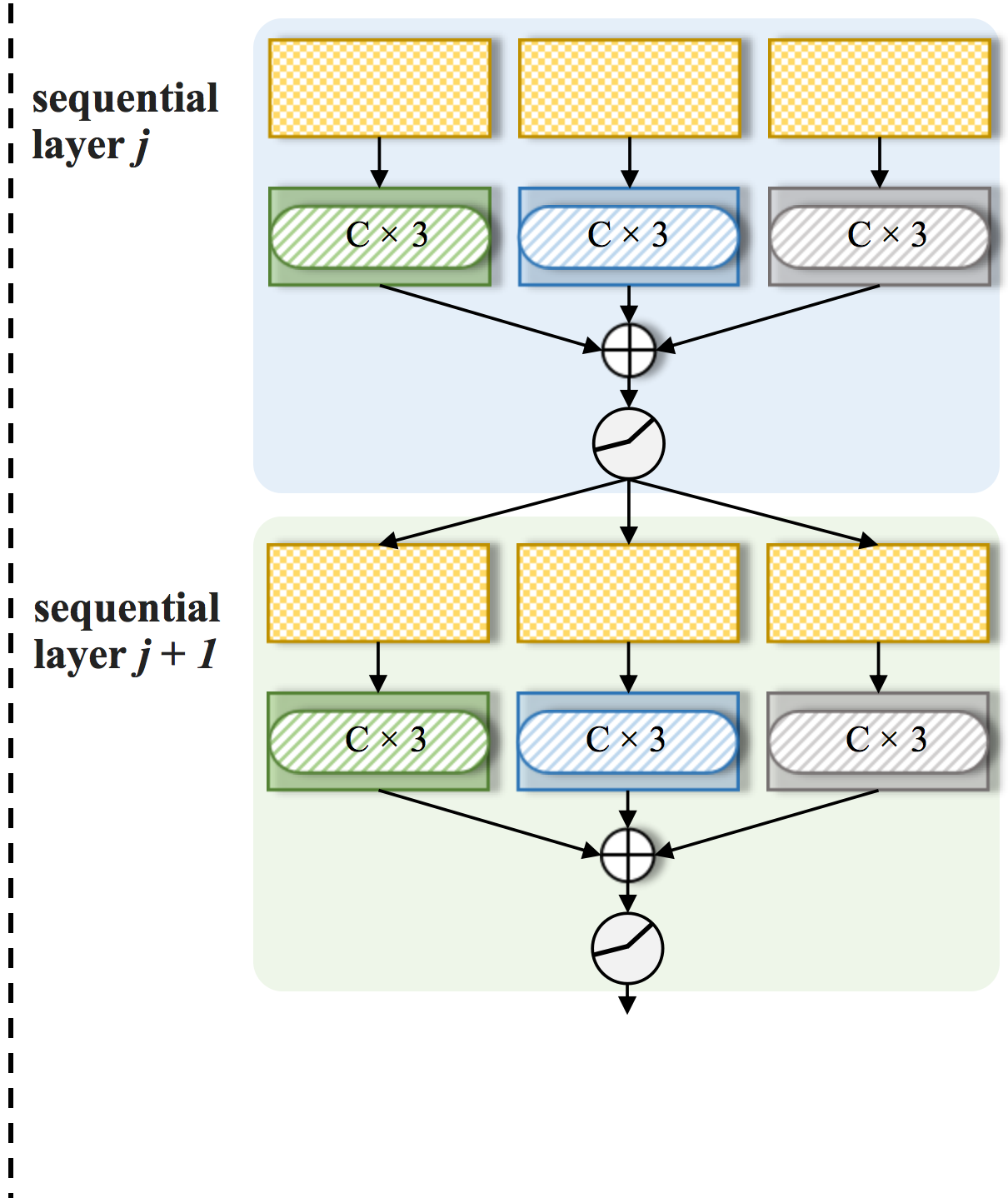, height=5cm}}
  \centerline{{(c)}}\medskip
\end{minipage}
\begin{minipage}[t]{0.24\linewidth}
  \centering
  \centerline{\epsfig{figure=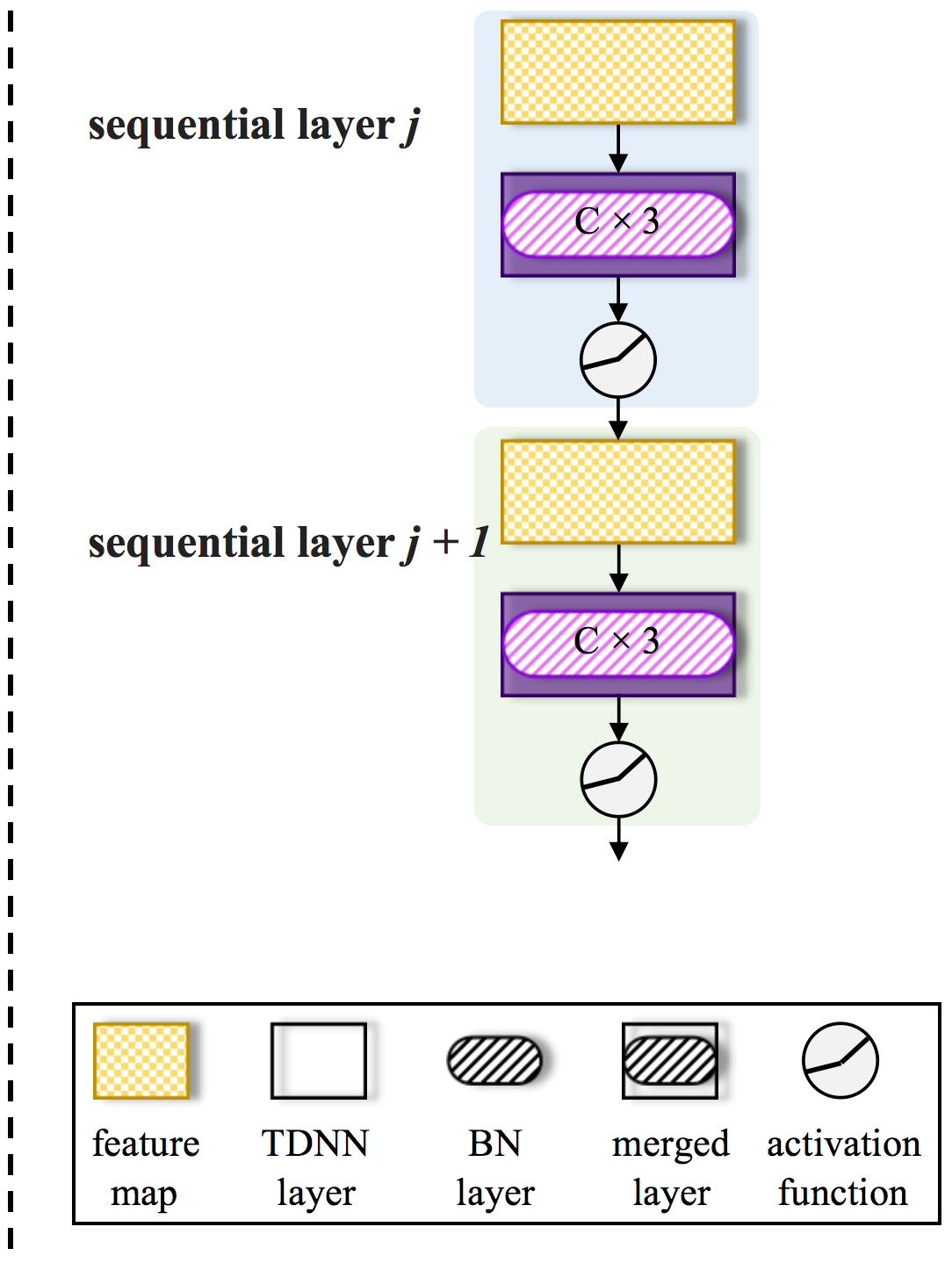, height=5cm}}
  \centerline{{(d)}}\medskip
\end{minipage}

\caption{\footnotesize{The topology changes for our model when adopting the proposed CS-Rep: (a) original topology of model, (b) step, (c) steps 2 and 3, and (d) step 4.}}
\label{cs_rep_step}
\vspace{-1em}

\end{figure*}

In this study, we propose the cross-sequential re-parameterization (CS-Rep, Fig. \ref{cs_rep_step}) to increase the inference speed and verification accuracy of ASV systems, especially for the models based on the ``conv-activation-bn”\footnote{The sequential layer \cite{paszke2019pytorch} is a layer including multiple modules by order. For example, the ``conv-activation-bn” is a sequential layer with the order of 1-D (TDNN) or 2-D convolutional layer, activation function, and BN layer. Fig. \ref{cs_rep_step} (a) shows the diagram of the sequential layer with three branches.} architecture. Specifically, we build a multi-branch network with the ``conv-activation-bn” structure during the training period. In the inference period, CS-Rep adjusts the order of modules of the sequential layer to the ``bn-conv-activation” losslessly, causing the BN to be adjacent to the convolutional layer. 
However, the ``bn-first"\footnote{The ``bn-first" means the case that the BN layer precedes the convolutional layer in the sequential layer. e.g., the ``bn-conv-activation.”} case will generate influences on each channel of the convolutional layer. We propose the ``bn-first" re-parameterization method to model these influences by the discrete convolution operator. 
Then, CS-Rep adopts this approach to convert the multi-branch network to a TDNN-like plain topology while maintaining the multi-scale information capturing ability. Therefore, the model adopted CS-Rep achieves an efficient architecture with friendly parallel computing.

The contributions of this study are listed as follows.
(1) We propose CS-Rep to increase the inference speed, decrease the parameters, and reduce the floating-point operations (FLOPs) of networks with the ``conv-activation-bn” structure while maintaining the accuracy. 
(2) We also combine the ``bn-first" and ``conv-first" \cite{ding2021repvgg} re-parameterization as an out-of-the-box method, which can be conveniently used for all types of sequential layers (``conv-first" and ``bn-first") to improve the inference speed.
(3) Based on CS-Rep, we construct Rep-TDNN. It achieves the state-of-the-art performance (1.09\% EER on VoxCeleb1-test, without any data augmentation), and its inference speed is 50\% faster than the state-of-the-art model ECAPA-TDNN\footnote{The code of ECAPA-TDNN is presented in \url{https://github.com/speechbrain/speechbrain}.}.

The remainder of the paper is organized as follows. Related works are briefly introduced in Section 2, the CS-Rep method and the novel Rep-TDNN model are detailed in Section 3, experimental results are presented and analyzed in Section 4, and the conclusions are given in Section 5.

\section{Related Work}

\noindent {\bf Time-delay neural networks}: The TDNN-based x-vector \cite{x-vector} system extracts robust embeddings by capturing the long-term temporal relations of utterances through the one-dimensional convolutional layer \cite{zhang2020aret}. Several improved versions of TDNN have been proposed, such as E-TDNN \cite{e-tdnn} with stable performance and satisfactory inference speed,  and ECAPA-TDNN \cite{ecapa-tdnn} with compelling accuracy.

\noindent {\bf Squeeze-excitation block}: Squeeze-excitation (SE) block \cite{hu2018squeeze} has been proved successful in modeling global channel interdependencies in speech, bringing  improved accuracy to ASV \cite{ecapa-tdnn}.

\noindent {\bf Multi-branch topology}: Multi-branch topology \cite{xie2017aggregated, gao2019res2net} involves multiple branches of different types in one sequential layer, so it helps to capture important information from multiple scales. In ASV, many solid models have adopted it, such as InceptionNet \cite{szegedy2017inception}, Res2Net \cite{gao2019res2net}, ECAPA-TDNN \cite{ecapa-tdnn}, and HS-Net \cite{zhang21_interspeech}. However,  multi-branch topology increases the complexity of network topology, leading to negative impacts on the inference speed.

\noindent {\bf RepVGG}: RepVGG \cite{ding2021repvgg} proposed a structural re-parameterization technique to improve the performance and reduce the computational complexity during inference. Their model applied the plain topology in the inference period and the multi-branch structure in the training period.
But there are two limitations in their method: (1) the re-parameterization method proposed in RepVGG is not suitable for ASV because the approach bases on the ``conv-bn-activation" structure, but mainstream ASV models adopt the ``conv-activation-bn" structure; (2) their work lacks the condition of the re-parameterization in TDNNs.

\section{Proposed methods}
\label{sec:proposed}

This section will describe the proposed CS-Rep strategy through a three-branch TDNN. CS-Rep will be used to develop Rep-TDNN with an extremely efficient architecture and high performance.

\subsection{Multi-branch training structure}
\label{bt_structure}
The three-branch structure \cite{ding2021repvgg} is introduced to build a TDNN block with excellent speaker information extraction. In our TDNN block, one sequential layer consists of three branches (Fig. \ref{cs_rep_step} (a)): a TDNN branch $TDNN^{(3)}$ with context = 3, a TDNN branch $TDNN^{(1)}$ with context = 1, and a shortcut connection, respectively. The operation within one sequential layer can be defined as:
\begin{eqnarray}
\label{multi_branch_formula}
\begin{aligned}
{\bf M}_{j}\!=\!B\!N\!_{j}(\!Re\!L\!U\!(T\!D\!N\!N\!^{(3)}_{j}(\!{\bf M}_{j\!-\!1}\!)\!+\!{T\!D\!N\!N\!}^{(1)}_{j}(\!{\bf M}_{j\!-\!1}\!)\!+\!{\bf M}_{j\!-\!1}\!)\!), 
\end{aligned}
\end{eqnarray}
where ${\bf M}_{j}\!\!\in\!\!\mathbb{R}^{B\!\times\!N\!\times\!T}$ is the output matrix of the ${j}$-th sequential layer, $B$ is the mini-batch size, $N$ is the number of the feature dimension, $T$ is the number of frames,  $ReLU$ is the ReLU function, and $B\!N_{j}$ is the batch normalization (BN) layer of the ${j}$-th sequential layer.

\subsection{Cross sequential re-parameterization method}
\label{sec:sc-rep}

This subsection describes how to convert a multi-branch network based on the ``conv-activation-bn" structure to a plain network using the cross sequential re-parameterization function (CS-Rep). Fig. \ref{cs_rep_step} and  Fig. \ref{bn-activation-conv}  display the topology diagram and parameter diagram for the network during the processing of CS-Rep.
The four steps of the approach are as follows:

{\bf Step 1: Cross-sequential converting}.
We distribute $BN_{j-1}$ from sequential-layer$_{j-1}$ to the head of the branches of sequential-layer$_{j}$, causing the model with the ``conv-activation-bn" structure to the ``bn-conv-activation" structure. This transform leads to the BN and TDNN layers contiguous and allowing them to be merged. This strategy is called cross-sequential transformation (CST).
The diagram of CST is displayed in Fig. \ref{cs_rep_step} (b).
After utilizing  CST, the sequential layer of Eq. (\ref{multi_branch_formula}) can be transformed to 
\begin{eqnarray}
\label{bn-activation-conv}
\begin{aligned}
{\bf M}_{j} = &ReLU({T\!D\!N\!N}^{(3)}_{j}(BN_{j-1}({\bf M}_{j-1})) + \\
&{T\!D\!N\!N}^{(1)}_{j}(BN_{j-1}({\bf M}_{j-1})) + BN_{j-1}({\bf M}_{j-1})).
\end{aligned}
\end{eqnarray}
Then we define that ${\bf W}^{(C)}_{j} \in \mathbb{R}^{N_{o} \times N_{i} \times C} $ is the weight of TDNN (${T\!D\!N\!N}_{j}$) from the $j$-th sequential layer,  $N_{i}$ and $N_{o}$ are the numbers of input and output channels respectively, $C$ is the number of the context of TDNN, and ``${*}$" is the discrete convolution operator. 
Eq. (\ref{bn-activation-conv}) can be re-written as:
\begin{eqnarray}
\begin{aligned}
{\bf M}_{j} = &ReLU({\bf W}^{(3)}_{j} * BN_{j-1}({\bf M}_{j-1}) +\\ 
&{\bf W}^{(1)}_{j} * BN_{j-1}({\bf M}_{j-1}) + BN_{j-1}({\bf M}_{j-1})).
\end{aligned}
\end{eqnarray}

\begin{figure}[t]
\setlength{\abovecaptionskip}{3pt}
 \centering
\includegraphics[width=1.0\columnwidth]{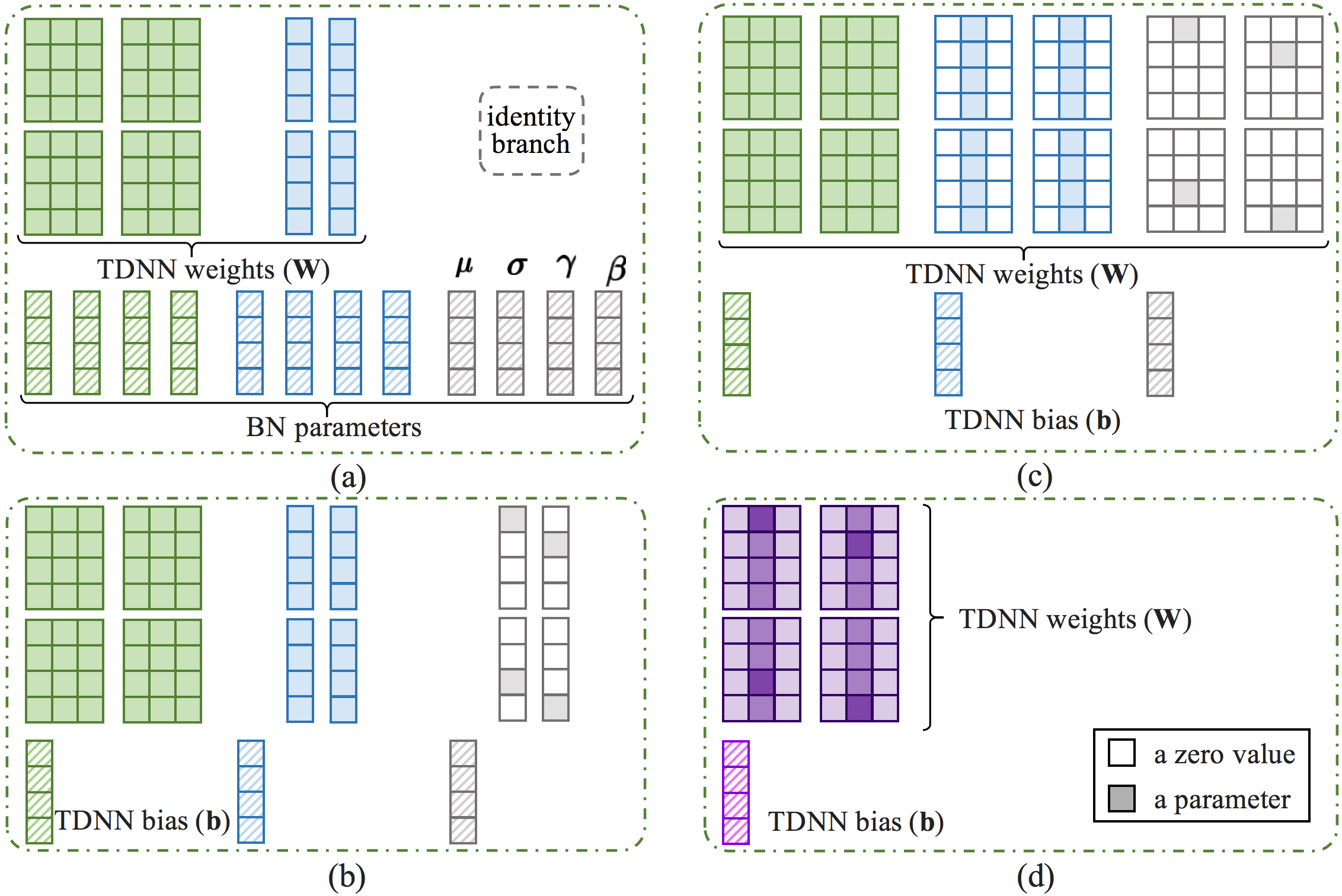}
\caption{\footnotesize{Parameters converting of the ``bn-first" re-parameterization in TDNN: (a) original parameters of weight and BN, (b) steps 1 and 2, (c) step 3, and (d) step 4.}}
\label{cs-rep-wieight}
\vspace{-1.5em}
\end{figure}

{\bf Step 2: Merge $BN_{j\!-\!1}$ into $TDNN_{j}$}. The BN \cite{ioffe2015batch}in TDNN is 
\begin{eqnarray}
\begin{aligned}
BN({\bf M}, \pmb{\mu}, \pmb{\sigma}, \pmb{\gamma}, \pmb{\beta})_{:,n_{o},:} = ({\bf M}_{:,n_{o},:} - \mu_{n_{o}}){\gamma_{n_{o}}\over{\sigma_{n_{o}}}} + \beta_{n_{o}}, 
\end{aligned}
\end{eqnarray}
where $n_o \in \{1,2,...,N_o\}$ denotes the $n_{o}$-th output channel, and $\pmb{\mu}$, $\pmb{\sigma}$, $\pmb{\gamma}$, and $\pmb{\beta}$ represent the mean of mini-batch, the variance of mini-batch, and the scale and shift parameters, respectively. And ${\mu}_{n_{o}}$, ${\sigma}_{n_{o}}$, ${\gamma}_{n_{o}}$, and ${\beta}_{n_{o}}$ are the elements of vectors $\pmb{\mu}$, $\pmb{\sigma}$, $\pmb{\gamma}$, and $\pmb{\beta}$, respectively.
In the ``bn-first" (BN is before the TDNN operator) re-parameterization, the condition that BN is before the convolutional layer will generate influences on each convolutional layer channel. 
The influence can be calculated as the bias of the convolutional layer through the discrete convolution operator.
The ``bn-first"  re-parameterization converts every TDNN branch and its preceding BN into a TDNN (${\bf W}$) with a bias vector (${\bf b}$) to reduce the depth of network (Fig. \ref{cs_rep_step} (c)), and it can be defined as:
\begin{eqnarray}
\begin{aligned}
\label{bn-first-rep}
&{\bf W}'_{j, {(n_{o},:,:)}}= {\bf W}_{j, {(n_o,:,:)}} \cdot  {{\gamma}_{j-1,(n_{o})}\over{{\sigma}_{j-1,(n_{o})}}}, \\
&{b}'_{j,(n_{o})}= {\bf W}_{j, {(n_o,:,:)}} * (-{\pmb{\gamma}_{j-1} \pmb{\mu}_{j-1}\over{\pmb{\sigma}_{j-1}}} + \pmb{\beta}_{j-1}), 
\end{aligned}
\end{eqnarray}
where $j$ denotes the $j$-th sequential layer, ${\bf W}'_{j}$ and ${\bf b}'_{j}$ are the weight and bias of the new TDNN branch through the ``bn-first" re-weight, and ${\bf W}'_{j, {(n_{o},:,:)}}$ is tensor slice of  ${\bf W}'_{j}$ and ${b}'_{j,(n_{o})}$ is the element of ${\bf b}'_{j}$.
Fig. \ref{cs-rep-wieight} (b) displays the weight diagram.

Then, we generalize a hot-swap algorithm for the ``bn-first" (Eq. (\ref{bn-first-rep})) and ``conv-first" \cite{ding2021repvgg}  re-parameterization methods.  
The algorithm can be written as: 
\begin{eqnarray}
\begin{aligned}
\label{bn-first_and_conv-first}
&{\bf W}'_{\!j, {(n_o,:,:)}}\!\!=\!\!{\bf W}_{\!j, {(n_o,:,:)}}\!\cdot\!\Phi_{n_{o}}; {\rm where} \ \pmb{\Phi}\!=\!\!\left\{
                \begin{array}{ll}
                  \!\!\!\!{\pmb{\gamma}_{j-1}\over{\pmb{\sigma}_{j-1}}},\!\!\!\!&{\rm \!bn \ first\!}\\
                  \!\!\!\!{\pmb{\gamma}_{j}\over{\pmb{\sigma}_{j}}},\!\!\!\!& {\rm \!conv \ first},
                \end{array}
              \right.\\
&{b}'_{j, (n_{o})\!}\!=\!\!\left\{
                \begin{array}{ll}
                  \!\!\!\!{\bf W}_{j, {(n_o,:,:)}}\!*\!(-{ \pmb {\gamma}_{j-1} \pmb{\mu}_{j-1}\over{\pmb{\sigma}_{j-1}}}\!+\!\pmb{\beta}_{j-1}\!),\!&{\rm \!bn \ first\!}\\
                  \!\!\!\!-{\!{\gamma}_{j,(n_{o})} {\mu}_{j,(n_{o})}\over {\sigma}_{j,(n_{o})}}\!+\!{\beta}_{j,(n_{o})},\!& {\rm \!conv \ first}.
                \end{array}
              \right.
\end{aligned}
\end{eqnarray}

For the re-weight of the identity branch, it can be converted to a TDNN branch with context = 1 (${\bf W}^{id}$):
\begin{eqnarray}
\begin{aligned}
\label{bn-first-rep-padding}
&\quad {\bf w}^{id}_{n_{o},n_{i},:}=\left\{
                \begin{array}{ll}
                  1, \quad n_{o}=n_{i}\\
                  0 ,\quad n_{o} \neq n_{i},
                \end{array}
              \right.
\end{aligned}
\end{eqnarray}
where ${\bf w}^{id}_{n_{o},n_{i},:}$ is the tensor slice of ${\bf W}^{id}$, $N_i$ and $N_o$ are the numbers of input and output channels of TDNN, and 
$n_i \in \{1,2,...,N_i\}$ and $n_o \in \{1,2,...,N_o\}$.
The diagram of building a TDNN layer equivalent to the shortcut connection is shown in the gray branch of Fig. \ref{cs-rep-wieight}(b). After converting the shortcut branch to the TDNN branch, it can be re-weighted by the ``bn-first” re-parameterization approach.

\begin{figure}[t]
\setlength{\abovecaptionskip}{2pt}
 \centering
\includegraphics[width=1.0\columnwidth,height=8cm]{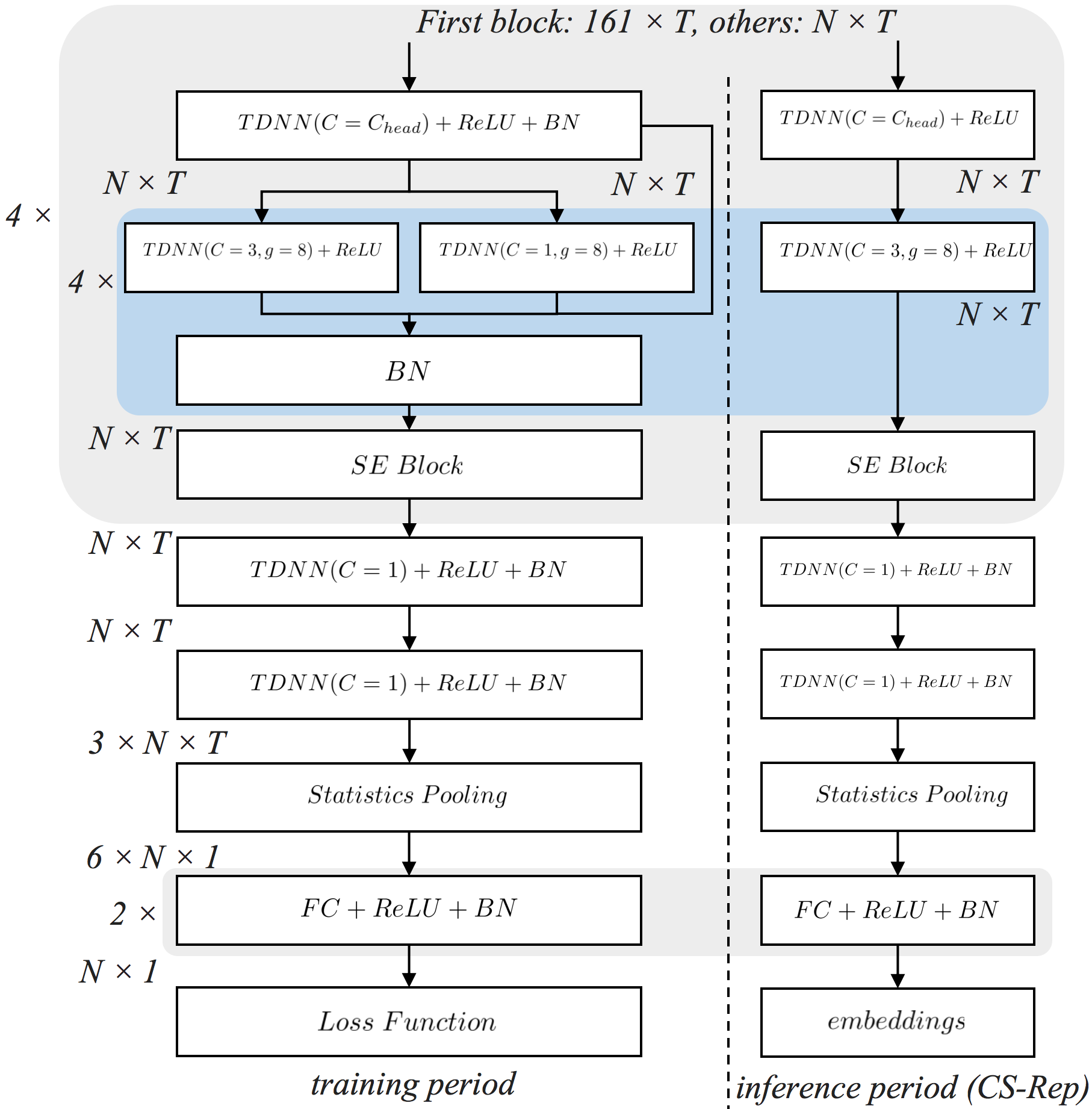}
\caption{\footnotesize{The topologies of Rep-TDNN without and with CS-Rep. The context size of the head TDNN in each block is $C_{head} \in [5, 1, 1, 5]$,  the $g$ represents the groups of layers, and the channel $N$ of each TDNN layer and branch is 512. \emph{FC}, \emph{BN}, \emph{SE} denote fully connected  layer, batch normalization, and squeeze-excitation, respectively. ({\bf Left}): The multi-branch design in the training period. ({\bf Right}): The plain topology in the inference period. }}
\label{Rep-ARET-structure}
\vspace{-1em}
\end{figure}

{\bf Step 3: Pad the weight of branches to the same dimension}. As for the TDNN branches with context = 1, we can use zero-padding on its weight to pad to context = 3 (Fig. \ref{cs-rep-wieight} (c)). 
After step 3 (Fig. \ref{cs_rep_step} (c)), the sequential-layer$_{j}$ is 
\begin{eqnarray}
\begin{aligned}
{\bf M}_{j}\!=&ReLU(({\bf W}'^{(3)}_{j}\!*\!{\bf M}_{j-1}\!+\!{\bf b}'^{(3)}_{j}) +\\
&(\!{\bf W}'^{(1\!\rightarrow\!3)\!}_{\!j\!}\!*\!{\bf M}_{j\!-\!1\!}\!+\!{\bf b}'^{(1\!\rightarrow\!3\!)}_{j}\!)\!+\!(\!{\bf W}'^{(0\!\rightarrow\!3)}_{j}\!*\!{\bf M}_{j\!-\!1\!}\!+\!{\!\bf b}'^{(0\!\rightarrow\!3)}_{j}\!)\!)\!,
\end{aligned}
\end{eqnarray}
where ${\bf W}'^{(1\rightarrow3)}_{j}$  means that the TDNN weight is changed from context = 1 to context = 3 through zero-padding. 
${\bf b}'_{j}$ is the bias of the corresponding branch, and the superscript of ${\bf b}'_{j}$ is only used to distinguish the branch uniformly while it does not mean padding.

\begin{table*}[t]
\setlength{\abovecaptionskip}{-0.00cm}
\setlength{\belowcaptionskip}{-1.5cm}
\renewcommand\tabcolsep{4pt}
\footnotesize
\begin{center}
  \caption{\footnotesize{The benchmark on VoxCeleb1 test, VoxCeleb1-E, and VoxCeleb1-H. Models were inferred on the single GPU to calculate the inference speed. Speed${\rm \textcolor{red}{^4}}$ in this table was defined as frame-numbers/processing-time.}}
  \label{results_all}
  \centering
  \begin{tabular}{ccccccccc}
    \toprule
     \multirow{2}{*}{\textbf{Description}}&   \multirow{2}{*}{\textbf{Backbone}}&   \multirow{2}{*}{\textbf{{Speed$\uparrow$}}}   &\multicolumn{2}{c}{\textbf{VoxCeleb1-test}} &\multicolumn{2}{c}{\textbf{VoxCeleb1-E}}  &\multicolumn{2}{c}{\textbf{VoxCeleb1-H}} \\
     &&&\textbf{EER$\downarrow$ (\%)}	&\textbf{minDCF$\downarrow$}&    \textbf{EER$\downarrow$ (\%)}	&\textbf{minDCF$\downarrow$}&	\textbf{EER$\downarrow$ (\%)} &\textbf{minDCF$\downarrow$}	\\
     \midrule
 	Our implementation &E-TDNN	(``conv-activation-bn”)			&175,512			&1.617 &0.2324			&1.584&0.3036		&2.771&0.4051 \\
	Our implementation &E-TDNN (``conv-bn-activation”)			&175,826			&2.096& 0.3579			&1.889&0.3629 	&3.358&0.4575 \\
	 \midrule
	Our implementation &ResNet-34							&72,656	&2.021&0.2696				&2.200&0.4136		&3.822&0.4753 \\
	Our implementation &ECAPA-TDNN							&62,802 			&1.181&0.2685				&1.343&0.2717 		&2.578&0.4188 \\
	 \midrule
	Proposed &Rep-TDNN (original topology)						&58,877			&1.117&0.1914				&1.193&0.2364		&2.163&0.3522 \\
	Proposed &Rep-TDNN (CS-Rep)							&92,903		&1.090&0.1964				&1.199&0.2377 		&2.272&0.3569 \\
    \bottomrule
  \end{tabular}
 \end{center}
\vspace{-2em}
\end{table*}

{\bf Step 4: Merge all branches to one TDNN layer}. Based on the linear additivity of the discrete convolution operator \cite{ding2021repvgg}, the weight and bias of every branch can be added to one branch to implement the merge-branch-convert easily. The final model only constructed by TDNN and ReLU(Fig.\ref{cs_rep_step}(d) and Fig.\ref{cs-rep-wieight}(d)),which can be written as:
\begin{eqnarray}
\begin{aligned}
&...\\
&{\bf M}_{j} = ReLU({T\!D\!N\!N}'^{(3)}_{j}({\bf M}_{j-1})) \\
&{\bf M}_{j+1} = ReLU({T\!D\!N\!N}'^{(3)}_{j+1}({\bf M}_{j}))\\
&...
\end{aligned}
\end{eqnarray}

\subsection{Proposed model}
Based on CS-Rep in Section \ref{sec:sc-rep}, we propose an improved TDNN called Rep-TDNN, which has excellent performance, fast inference speed, few parameters, and fewer floating-point operations (FLOPs). The overview of Rep-TDNN is displayed in Fig. \ref{Rep-ARET-structure}. The frame-level model consists of four blocks, including TDNN layers, LeakyReLU activation functions, BN, and SE-Block.

In the training period (Fig. \ref{Rep-ARET-structure} (right)), each block has one head TDNN layer (the first TDNN layer in the gray box) that integrates feature information and four TDNN sequential layers with three branches (a branch with context = 3, a branch  with context = 1, and a shortcut branch; describes in Section \ref{bt_structure}) with the ``conv-activation-bn" structure. The context $C_{head}$ of each head TDNN layer is [5, 1, 1, 5]. SE-block is adopted in each block to model the channel attention. Statistics pooling aggregates frame-level features to segment-level features passed to two fully connected (FC) layers. AAM-Softmax \cite{arc-softmax} loss is to choose for loss function. \blfootnote{${\rm ^4}$Other common metrics for speed, e.g., RTF, RTX, FPS, etc., which can be easily converted from these results.}

In the inference period, the CS-Rep method is adopted to convert the model from the multi-branch topology to the plain topology (Fig. \ref{Rep-ARET-structure} (left)). 
The parameters and FLOPs of Rep-TDNN (CS-Rep) are only 6.9e+6 and 1.4e+9, which are the same as for E-TDNN \cite{ecapa-tdnn, zhang2020aret}.

\section{experiments}
\subsection{Experimental details}
\noindent {\bf Dataset}: To investigate the performance of the models, we created a benchmark on the VoxCeleb \cite{vox1,vox2} dataset.
We implemented some successful architectures, including E-TDNN \cite{zhang2020aret}, ResNet34 \cite{ecapa-tdnn}, and ECAPA-TDNN \cite{ecapa-tdnn}, as our baselines.
Models were trained on the VoxCeleb2 development sets without data augmentation, and tested on the VoxCeleb1 test set, VoxCeleb1-E, and VoxCeleb1-H \cite{vox2}. 
161-dimensional spectrograms were generated through hamming sliding window with a width of 20 ms and a step of 10 ms. The cepstral mean and variance normalization (CMVN) was applied. No voice activity detection was adopted.

\noindent {\bf Implementation details}:
Models were trained by AAM-Softmax loss with $m=0.25$ and $s=30$. In the training stage, the mini-batch size of 64 was used. Each training speech sample was randomly sampled for 300 frames from recordings. Stochastic gradient descent (SGD) with momentum 0.9, weight decay 1e-5, and initial learning rate 0.1 was utilized. Full-length utterances were adopted in the testing stage to extract embeddings. The GPU type and CUDA version were NVIDIA 2080Ti and 10.2.
Adaptive score normalization (AS-Norm) \cite{cumani11_interspeech} with cosine similarity was applied for all models. L2-normalized speaker embeddings of each training speaker were chosen as the imposter cohort with a size of 1000.  

\noindent {\bf Evaluation metrics}: We utilized the equal error rate (EER) and minimum detection cost function (minDCF) from NIST SRE10 \cite{sre10} as the performance metrics. 
Moreover, we adopted the actual inference speed (frame-numbers/processing-time) to compare the speed of different models. It is more objective than the indirect speed metrics parameters and FLOPs, which can avoid the bad case in which a model with few parameters and complex topology has a low FLOPs but the actual inference speed may be slow \cite{ma2018shufflenet}. 
\subsection{Experimental results}
\label{sec:exp_results}

The results of the state-of-the-art baseline on VoxCeleb1-test, VoxCeleb1-E, and VoxCeleb1-H are shown in Table \ref{results_all}. The results of the inference speed are the average of 5 runs.

In Table \ref{results_all}, the E-TDNN with the ``conv-activation-bn” structure obtained approximately 1.6\%, 1.6\%, and 2.8\% EER on the above three evaluation sets, respectively, whereas the E-TDNN with the ``conv-bn-activation” structure only achieved 2.1\%, 1.9\%, and 3.4\% EER, respectively. 
These results confirm that the ``conv-activation-bn” architecture is necessary for high-accuracy ASV networks.

Compared with the best E-TDNN, the EER results of Rep-TDNN were much lower, relatively reduced by 32.6\%, 24.3\%, and 18.0\% on three evaluation conditions, respectively. Satisfactory performances showed the strong discrimination of speakers for Rep-TDNN.
As shown in Table \ref{results_all}, Rep-TDNN obtained approximately 1.2\% EER and 2.2\% EER on VoxCeleb1-E and VoxCeleb1-H, respectively, which are both hard evaluation test sets.
Compared with the results with ECAPA-TDNN, Rep-TDNN had 10.7\% and 11.9\% reductions of EER on two hard test sets and a 47.9\% increase of inference speed.
We can find that compared to ECAPA-TDNN, Rep-TDNN was more efficient and more robust in hard conditions.

Furthermore, CS-Rep can increase the inference speed for the model with negligible influence on the accuracy. 
Compared with the Rep-TDNN without CS-Rep, the inference speed of the model with CS-Rep was greatly improved (by 58\%) while EERs stayed the same. In Table \ref{results_all}, Rep-TDNN through CS-Rep can process 92,903 frames per second, whereas Rep-TDNN with original topology only can process 58,877 frames per second. The results show that CS-Rep is an essential component for developing ASV models with the fast inference speed and high accuracy.

Generally, Rep-TDNN showed relatively improvements of approximately 50\% in inference speed and 10\% in performance compared with state-of-the-art ASV models. Hence, it has great  potential for the industry, saving spends for test and deployment.

\section{conclusion}

This study presents an upgraded re-parameterization approach called CS-Rep to re-parameterize models with a multi-branch design more efficiently. It can be used to reduce the size of parameters, reduce inference time, and improve the performance of ASV systems. We also integrate the ``bn-first" and ``conv-first" of re-parameterization equations into a plug-and-play module. 
Based on CS-Rep, this work proposes Rep-TDNN, which obtains 18\% -- 32.6\% relative reduction in EER compared with solid baselines. 
The positive results show that Rep-TDNN reduces the inference time by approximately 50\% and improves the performance by approximately 10\% compared with the state-of-the-art model  ECAPA-TDNN.
%





\bibliographystyle{IEEEbib}
\bibliography{strings,refs}

\end{document}